\documentstyle[twocolumn,seceq,epsf]{jpsj2}
\newcommand{\plcco}{Pr$_{1-x}$LaCe$_x$CuO$_{4+\alpha-\delta}$}
\newcommand{\plccod}{Pr$_{0.89}$LaCe$_{0.11}$CuO$_{4+\alpha-\delta}$}
\newcommand{\msr}{$\mu$SR}

\title
{Magnetic Ground State of \plccod\ with Varied Oxygen Depletion
Probed by Muon Spin Relaxation
}

\author
{ 
Ryosuke {\sc Kadono}\thanks{Also at School of Mathematical and Physical Science, 
The Graduate University for Advanced Studies}, 
Kazuki {\sc Ohishi}, Akihiro {\sc Koda}, Wataru {\sc Higemoto}, 
Kenji M. {\sc Kojima}$^1$\thanks{Present Address: Department of Physics,
University of Tokyo, Tokyo 113-8656}\\
Shin-ichi {\sc Kuroshima}$^2$, Masaki {\sc Fujita}$^2$, 
and Kazyoshi {\sc Yamada}$^2$\thanks{Present Address: Institute for Materials
Research, Tohoku University, Sendai 980-8577}
}

\inst
{
Institute of Materials Structure Science, High Energy Accelerator Research Organization (KEK),
Tsukuba, Ibaraki 305-0801\\
$^1$Graduate School of Frontier Sciences , University of Tokyo, Tokyo 113-8656\\
$^2$Institute for Chemical Research, Kyoto University, Uji, Kyoto 610-0011 
}

\recdate
{\hspace{1cm}
}

\abst
{
The magnetic ground state of an electron-doped cuprate
superconductor 
\plcco\ ($x=0.11,\; \alpha\simeq0.04$) has been studied by 
means of muon spin rotation/relaxation (\msr)
over a wide variety of oxygen depletion, $0.03\le\delta\le0.12$. 
Appearance of weak random magnetism over entire crystal 
volume has been revealed by a slow exponential relaxation. 
The absence of $\delta$-dependence for the random magnetism and
the multiplet pattern of muon Knight shift at higher fields
strongly suggest that the random moments are associated with
excited Pr$^{3+}$ ions under crystal electric field.
}
\kword
{
superconductivity, crystal electric field, cuprates, \msr
}

\begin{document}
\sloppy
\maketitle

It is widely believed that superconductivity of electron-doped
cuprates is in an intimate relationship with that of hole-doped
cuprates due to the common background of 
CuO$_2$ planes.\cite{Tokura:89,Takagi:89} 
In this regard, electron-hole asymmetry of the phase diagram 
observed between those two groups\cite{Takagi:89,Luke:89} is one of the key issues 
for selecting the models of pairing mechanisms being proposed.
Despite its importance, however, so far the study of electron-doped
cuprates is far behind that of hole-doped systems.
This is partly because of the limited variety of 
compounds and associated difficulty to obtain them as large
single crystal specimen; earlier works have been done
mostly on $L_{2-x}$Ce$_x$CuO$_{4-\delta}$
(where $L$=Nd, Pr, and Sm) which are often available 
only as a small amount of tiny crystals.
This situation has been changed recently by the 
development of a new electron-doped superconductor
Pr$_{1-x}$LaCe$_x$CuO$_{4-\delta}$ (PLCCO) and 
subsequent success in growing a large single crystal.\cite{Fujita:03,Isawa:02}

Another important issue related to the electron-doped
cuprates is the microscopic nature of the oxygen depletion 
process which is needed to turn the as-grown material into
superconducting.\cite{Tokura:89,Takagi:89}
It has been presumed that the oxygen atoms in the CuO$_2$
planes become slightly deficient upon depletion procedure.  In this situation,
the mobility of carriers may be affected by the introduced randomness
in CuO$_2$ planes, while the effect on the doping level would be simply a
positive shift.\cite{Tokura:89,Takagi:89} Another possibility is that 
there are excess oxygen atoms
at apical sites between the CuO$_2$ planes which are unoccupied
in the final T' structure after depletion procedure.
The structure having apical oxygen resembles
that of T$^*$ phase in 
Nd$_{2-x-z}$Ce$_x$Sr$_z$CuO$_{4-\delta}$\cite{Akimitsu:88,Sawa:89}
whose charge carriers are identified as holes,
thus suggesting that a carrier compensation process
similar to the case in semiconductors occurs in the as-grown
material. The newly synthesized
PLCCO has an interesting character that it is stable over a
wide range of the oxygen depletion, $0\le\delta\le0.12$. 
This makes it feasible to investigate the effect of oxygen
depletion on CuO$_2$ planes in more detail.

In this Letter, we report on our muon spin rotation/relaxation (\msr) experiment
in an electron-doped superconductor \plccod\ ($\alpha\simeq0.04$) to elucidate the 
ground state phase diagram with $\delta$ as the primary parameter.
(The phase diagram as a function of carrier concentration $x$
has been reported elsewhere,\cite{Fujita:03} yielding a result similar to 
Nd$_{2-x}$Ce$_x$CuO$_{4-\delta}$ (NCCO) with a wider region
of superconductivity over $x$ from 0.09 to 0.20.)
The specimen exhibits a maximum of 
superconducting transition temperature ($T_c$)  around $\delta=0.06$,
indicating that the oxygen depletion has a strong influence on the doping 
level with a partial compensation of carriers for $\delta<0.06$.
This is consistent with the presence of excess oxygen at apical sites.
On the other hand, the \msr\ spectra under zero field 
exhibit little dependence on $\delta$, with a common tendency 
of slow exponential depolarization developing with decreasing temperature
below $\sim150$ K.  This indicates the presence of random
local magnetic moments irrespective of oxygen depletion.
We also found that the muon Knight shift exhibits a triplet
structure at high magnetic field.   These results strongly suggest
that the weak random magnetism is primarily due to Pr$^{3+}$ ions,
whereas that associated with Cu$^{2+}$ spins appears only
for $\delta\le0.03$.

\begin{figure}
%\figureheight{11cm}
\begin{center}
\mbox{\epsfxsize=0.4\textwidth \epsfbox{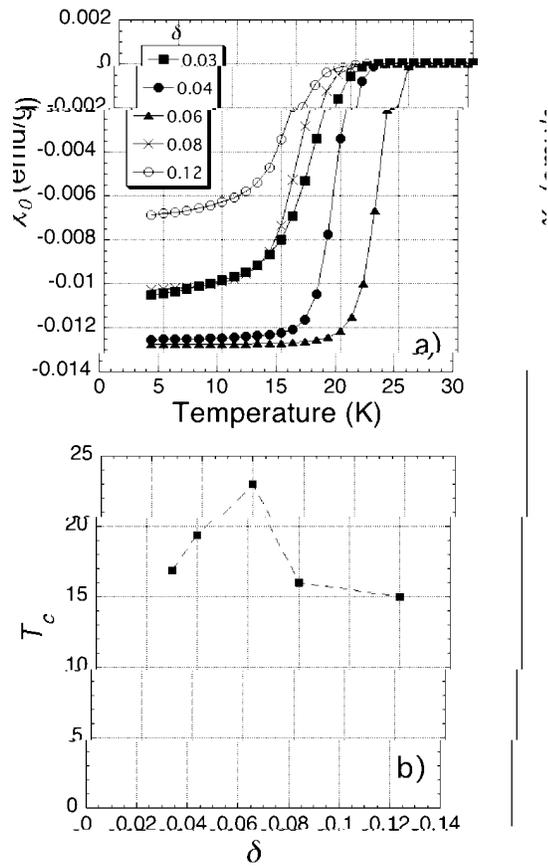}}
\end{center}
\caption{a) Temperature dependence of the magnetic ac-susceptibility
$\chi_0$ in \plccod, where $\chi_0\simeq$0.013 emu/g corresponds
to $1/4\pi$ (fractional yield of 100 \% for the Meissener effect).
b) Superconducting transition temperature $T_c$ vs degree of oxygen depletion
$\delta$, where $T_c$ is defined as a mid-point of the Meissener fraction.
}\label{chi-ac}
\end{figure}

A single crystal of Pr$_{0.89}$LaCe$_{0.11}$CuO$_{4+\alpha}$
were prepared by a traveling-solvent floating zone
method, where the detail of the procedure is reported
previously.\cite{Fujita:03} 
The presence of excess oxygen in the as-grown crystal
has been confirmed by an iodometric titration 
technique to yield $\alpha\simeq0.04$.
The crystal was sliced
into slabs measuring about 5 mm $\times$ 8 mm $\times$
0.5 mm thick with $c$-axis being perpendicular to the plane, 
which were then annealed under 
argon gas flow to reduce oxygen to the respective
level. The amount of removed oxygen $\delta$ per unit formula
from as-grown specimen was determined by the weight
loss after the annealing treatment.  For the present experiment
we obtained specimen with $\delta$=0.03, 0.04, 0.06, 0.08, and 0.12.
The result of magnetic ac-susceptibility (1 mT, 100 Hz) 
 is shown in Fig.~\ref{chi-ac}a,
from which $T_c$ is determined as a mid-point of the Meissener
effect (see Fig.~\ref{chi-ac}b). It is noteworthy in those 
figures that the optimal ``doping" occurs when $\delta\simeq0.06$,
with a clear tendency of decresing $T_c$ and Meissener fraction
with decreasing $\delta$.  This behavior is reminiscent to the case
of hole-doped cuprates in the underdoped region, which is in marked
contrast with their monotonical dependence on the Ce content $x$.\cite{Fujita:03}
A similar dependence of $T_c$ on the oxygen depletion has been reported for 
NCCO ($x=0.15$).\cite{Kurahashi:02}

\begin{figure}
%\figureheight{11cm}
\begin{center}
\mbox{\epsfxsize=0.4\textwidth \epsfbox{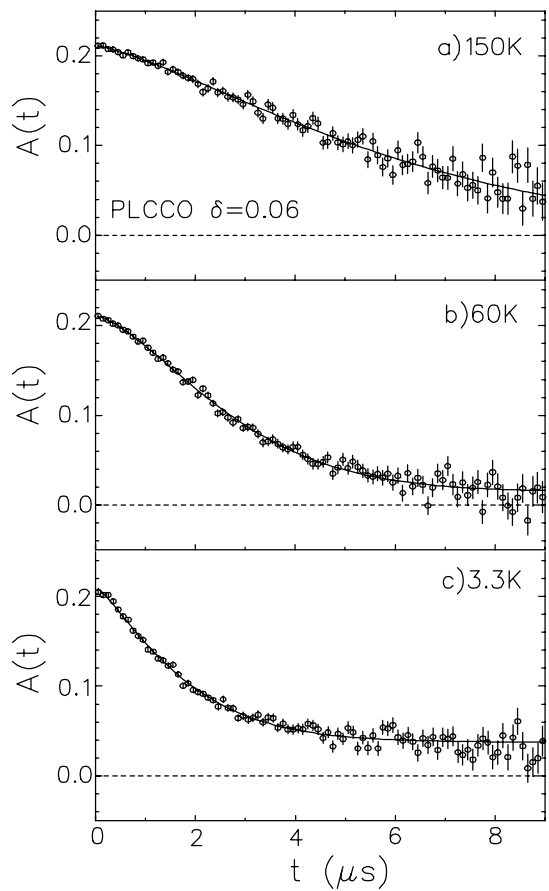}}
\end{center}
\caption{ZF-\msr\ time spectra observed in \plccod\ with $\delta=0.06$.  
An exponential depolarization gradually takes over the Gaussian damping
due to random local fields from nuclear magnetic moments.
Those in other samples are quite similar to these examples.
}\label{tspec}
\end{figure}

Conventional \msr\ 
measurements on those samples under zero and transverse external field 
were performed on the M15 beamline of TRIUMF.
The observed ZF-\msr\ time spectra in those specimens
are common to show an exponential depolarization overlapped
with Gaussian-like damping due to nuclear dipolar
fields, indicating that the origin of the exponential relaxation 
is due to random magnetic moments.  Some examples from the
data with $\delta=0.06$ are
shown in Fig.~\ref{tspec}, where the exponential damping has
almost full fraction of the positron decay asymmetry at 3.3 K.
Note that the Meissener fraction is almost 100 \% in this specimen.
This indicates that the entire volume of the specimen is subject
to the weak random magnetism irrespective of superconductivity.
The time spectra under a longitudinal field ($\simeq15$ mT)
do not exhibit appreciable relaxation at 40 K, indicating that the
random fields are nearly static within the time window
of \msr\ ($\sim10^{-5}$ s) at lower temperatures.

\begin{figure}
%\figureheight{11cm}
\begin{center}
\mbox{\epsfxsize=0.4\textwidth \epsfbox{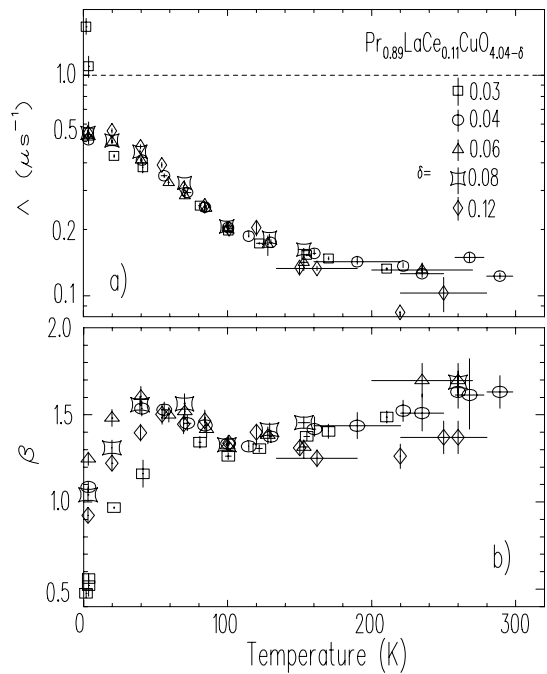}}
\end{center}
\caption{Temperature dependence of muon spin relaxation rate
$\Lambda$ (a) and associated power $\beta$ (b)
under zero external field observed in \plccod.  
A steep increase of $\Lambda$ is seen only in the specimen
with $\delta=0.03$ below $\sim$2 K.
}\label{rlxrate}
\end{figure}

For a quantitative analysis, the ZF-\msr\ spectra were fit by the
stretched exponential damping;
\begin{equation}
A(t)=A_0\exp[-(\Lambda t)^\beta]+B,
\end{equation}
where $A_0$ is the initial positron decay asymmetry
(which is proportional to the muon polarization), $\Lambda$
is the spin relaxation rate, and $\beta$ is the power of
the damping.
The results of fitting analysis by the above equation 
are summarized in Fig.~\ref{rlxrate}.
The relaxation rate $\Lambda$ exhibits a universal behavior
of increase with decreasing temperature below
about 150 K regardless of the oxygen depletion,
except that below 2 K in the specimen with $\delta=0.03$
where a steep increase of $\Lambda$ is observed.
The latter behavior ($\delta=0.03$, below 2 K)
is in line with a spin glass-like magnetism suggested by
the recent result of magnetization measurement (see below).\cite{Kuroshima:pr}
Using a coarse approximation,
$\Lambda\simeq \gamma_\mu[\langle H_n^2\rangle+
\langle H_e^2\rangle]^{1/2}$ (with $\gamma_\mu$ being the muon gyromagnetic
ratio, $\langle H_n^2\rangle$ and $\langle H_e^2\rangle$ being the  
variance of respective local fields from nuclei and magnetic ions),
one can estimate the contribution of magnetic ions.
Since the leveling off of $\Lambda$ above 150 K is mainly attributed
to the contribution of nuclear random local fields (i.e., 
$\Lambda\simeq\gamma_\mu\langle H_n^2\rangle^{1/2}\sim0.1$ 
$\mu$s$^{-1}$), 
the contribution of magnetic ions can be estimated 
to be $\langle H_e^2\rangle^{1/2}\simeq 0.47$ mT.
The small power below $\sim50$ K suggests slowing down of the
fluctuating random fields, while the similar behavior
at higher temperatures can be understood by considering diffusive
motion of muons.

\begin{figure}
%\figureheight{11cm}
\begin{center}
\mbox{\epsfxsize=0.4\textwidth \epsfbox{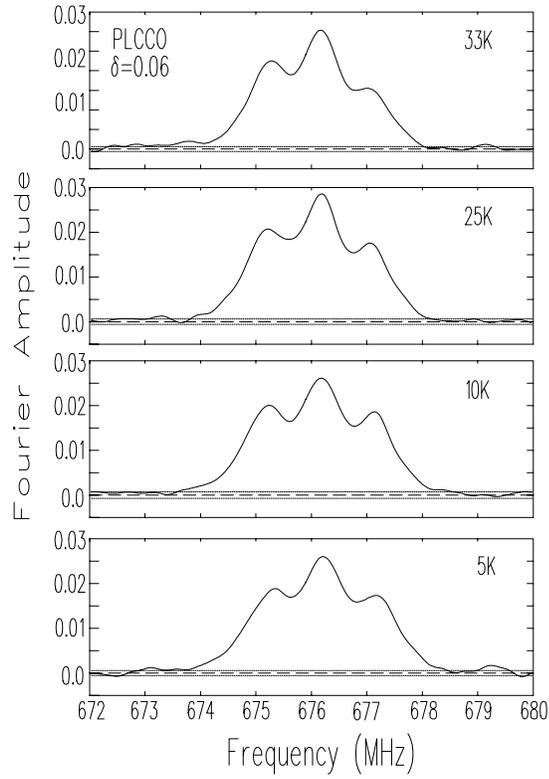}}
\end{center}
\caption{Fast Fourier transform of \msr\ time spectra measured
under a transverse field of 5 T in \plccod\ with $\delta=0.06$. 
The peak at the highest frequency ($\sim677.2$ MHz) 
corresponds to the component
with null Knight shift. 
}\label{fft}
\end{figure}

It has been revealed in the very recent study that
the spin-glass phase observed near the antiferromagnetic (AF)
phase of \plcco\ for $\delta\le0.03$ strongly depends on both Ce concentration
$x$ and oxgen depletion $\delta$,\cite{Kuroshima:pr}
where the magnetic moments are apparently carried by copper spins.
Thus, the absence of $\delta$ dependence in the present result suggests
that the random moments observed for $\delta>0.03$ 
are carried by Pr$^{3+}$ ions, although the ground state of which is presumed
to be nonmagnetic.  This is further confirmed by the
structure of muon Knight shift under high magnetic
fields.  As shown in Fig.~\ref{fft}, the Fourier transform
of \msr\ time spectra splits into three peeks
with a relative amplitude of 1:2:1.  This can be readily
understood by considering the situation that there
are two nearest neighboring ions contributing to the
Knight shift with different combination of Pr$^{3+}$
and La$^{3+}$ (neglecting minor contribution of Ce$^{3+}$).
The shift under 5 T
is about 0.15 \% for the central peak and 0.3 \% 
for the lowest frequency peak at 5 K, which is least dependent on temperature.
These results indicate that there is a
significant contribution of magnetic excited levels
split by the crystal electric field over the relevant
temperature range, leading to the Van Vleck 
magnetism at higher magnetic fields.
A similar situation has been observed in 
Pr$_{1.85}$Ce$_{0.15}$CuO$_{4-\delta}$ (PCCO).\cite{Sonier:00}

Provided that implanted muons occupy the sites near the edge center 
of tetrahedron with Pr/La atoms at their corners, one can estimate
the magnitude of effective dipolar fields at the muon sites by
calculating a dipolar tensor 
$A^{\alpha\beta}_i=(\delta_{\alpha\beta}-3r_i^\alpha r_i^\beta/r_i^2)/r_i^3$ 
and the relevant variance 
$(\overline{A}_{xy})^2=\sum_{i,\alpha,\beta}(A^{\alpha\beta}_i)^2$
from $i$-th ion at a distance ${\bf r}_i$ ($\alpha=x,y$ and $\beta=x,y,z$ when the
primary axis is set to $z$). 
Taking account of the nearest two ions, we obtain
$\langle H_e^2\rangle^{1/2}\simeq\overline{A}^{\rm Pr}_{xy}\simeq0.19$ 
T/$\mu_B$ (unit Bohr magneton) as an average of all possible combinations
for Pr and La ions.
The comparison of the above estimation with the observed 
magnitude of $\langle H_e^2\rangle^{1/2}$ ($\sim0.47$ mT)
implies that the Pr$^{3+}$ ions have a magnetic moment
\begin{equation} 
|\mu_{\rm Pr}|=\frac{\langle H_e^2\rangle^{1/2}}{\overline{A}^{\rm Pr}_{xy}}
\simeq2.4\times10^{-3}\mu_B
\end{equation}
under zero external field.  This is more than by an order of magnitude 
smaller than that observed in the AF phase of 
Pr$_2$CuO$_4$ where $|\mu_{\rm Pr}|=0.08\mu_B$.\cite{Matsuda:90}

The Pr$^{3+}$ free-ion $^3H_4$ state multiplet
has ninefold degeneracy, which splits into 
five singlets (2$\Gamma_1$, $\Gamma_2$, $\Gamma_3$, $\Gamma_4$)
and two magnetic doublets (2$\Gamma_5$) under
the tetragonal $C_{4v}$ symmetry of PLCCO.
Although we do not have direct information on the Pr$^{3+}$ ions
in PLCCO at this stage,
there are several literatures on the crystal field effects in 
Pr$_2$CuO$_4$ and PCCO studied by 
inelastic neutron scattering\cite{Boothroyd:92,Loong:93} 
and Raman scattering,\cite{Sanjurjo:95,Jandl:97}
where the first excited state is reported to be a $\Gamma_5$ doublet
which is separated from a singlet (either $\Gamma_3$ or
$\Gamma_4$) ground state by 18 meV.  
Considering the nearly identical T' structure, 
this situation can be presumed to be the
case also for the Pr$^{3+}$ ions in PLCCO.
Then, the observed weak magnetic moment can be
attributed to the small mixing of $\Gamma_5$ state
with the singlet ground state, whereas the moment is
considerably enhanced in the AF phase of Pr$_2$CuO$_4$ due
to the Van Vleck magnetism.

Interestingly, it happens that the muon spin relaxation
rate in NCCO ($x=0.15$) exhibits
a similar tendency of gradual increase with decreasing
temperature except below $\sim2$ K where a steep
increase sets in.\cite{Luke:97} Assuming that the Nd
moments are quasi-static below 2 K, the observed relaxation
rate ($>1$ $\mu$s$^{-1}$) suggests that the Nd$^{3+}$
ions have a moment considerably larger than that of Pr$^{3+}$.
In both cases, since the sample crystal is a good superconductor as
inferred from the large fraction of Meissener diamagnetism, we 
can conclude that the magnetic moments of
the rare earth ions do not directly interfere the superconductivity on
the CuO$_2$ planes.

Given that the weak random magnetism is entirely due to 
Pr$^{3+}$ ions in \plccod, it also means that there is no appreciable
contribution of copper spins, if at all, over the entire
region of $\delta$ (except below $\sim$2 K in the specimen with 
$\delta=0.03$).  
According to the earlier neutron diffraction study, the Cu$^{2+}$ spins
have a moment $\simeq0.4\mu_B$ in Pr$_2$CuO$_4$,\cite{Matsuda:90}
 in which muons feel an internal field  of $\sim27$ mT.\cite{Luke:89}
Assuming the same site for muons as discussed previously, 
the calculated dipolar tensor for the  Cu$^{2+}$ ions
yields $\overline{A}^{\rm Cu}_{xy}\simeq 50.2$ mT/$\mu_B$ and thus
qualitatively consistent with the earlier result of \msr.
Since PLCCO is nearly isostructural to Pr$_2$CuO$_4$, these estimations indicate
that the present measurement must be sensitive to the
presence of quasi-static copper spins in the order of $10^{-2}\mu_B$.
Taking the contribution of {\sl nuclear} dipolar fields ($\Lambda\simeq0.1$ 
$\mu$s$^{-1}$) as a limiting background, 
we can place an upper bound for the quasi-static copper moment;
\begin{equation}
|\mu_{\rm Cu}|_{\rm static}<0.015\mu_B.
\end{equation}
It must be noted, however, that this does not exclude the presence of
copper moments which are fluctuating with a time scale shorter than
that susceptible to \msr\ ($<10^{-9}$ s).

Finally, we discuss the chemistry of oxygen depletion in the present
PLCCO system.  As shown in Fig.~\ref{chi-ac}, the Meissener
effect is at its maximum when $\delta=0.04\sim0.06$, which
is in good agreement with the amount of excess oxygen ($\alpha\simeq0.04$).
When the oxygen depletion proceeds, the superconducting property is
rapidly deteriorated as indicated by the marked decrease
of both Meissener fraction and $T_c$.
This observation strongly suggests that the optimal
superconductivity is realized when oxygen atoms are
in the stoichiometric composition, Pr$_{2-x}$LaCe$_x$CuO$_4$.
It is likely that as-grown crystals always have some excess oxygen atoms 
at the apical sites, which serves as defect centers giving rise to carrier compensation.
While the removal of the excess oxygen improve the superconducting
property of PLCCO, it seems that further removal of oxygen erodes the
CuO$_2$ planes, leading to the rapid deterioration of superconductivity.

In summary, we have demonstrated that, while the bulk superconducting 
property of single crystalline \plccod\ ($\alpha\simeq0.04$)
exhibits a considerable dependence on the oxygen depletion
$\delta$ over the region studied ($0.03\le\delta\le0.12$),
the magnetic ground state probed by \msr\, which is characterized by a weak random
magnetism, is least dependent on $\delta$. 
The muon Knight shift strongly suggests that the random
magnetism is due to a small magnetic moment of Pr$^{3+}$
ions induced by the mixing of an excited state under crystal
electric field.  Based on these observations, a consistent microscopic 
view of the oxygen depletion has been provided by the present study.

We would like to thank the staff of TRIUMF for their technical
support during the experiment. We also appreciate discussion
with T. Uefuji on the earlier result of NCCO.
This work was partially supported by a
Grant-in-Aid for Scientific Research on Priority Areas and
a Grant-in-Aid for Creative Scientific Research from the Ministry of
Education, Culture, Sports, Science and Technology of Japan.

%\newpage

%Figure Captions
\newpage

\newpage

\newpage

\newpage

\end{document}